\documentclass[12pt]{article}
\usepackage{graphicx}
\usepackage{amsmath}
\usepackage{longtable}

\begin{document}

\makeatletter
\renewcommand*{\@cite}[2]{{#2}}
\renewcommand*{\@biblabel}[1]{#1.\hfill}
\makeatother

\title{Three-Dimensional Reddening Map for Stars from 2MASS Photometry: The Method and the First Results}
\author{G.~A.~Gontcharov\thanks{E-mail: georgegontcharov@yahoo.com}}

\maketitle

Pulkovo Astronomical Observatory, Russian Academy of Sciences, Pul\-kov\-skoe sh. 65, St. Petersburg, 196140 Russia


The first results of the construction of a three-dimensional reddening map for stars within
1600 pc of the Sun are presented. Analysis of the distribution of 70 million stars from the 2MASS
catalog with the most accurate photometry on the $(J-Ks)$ -- $Ks$ diagram supplemented with Monte
Carlo simulations has shown that one of the maxima of this distribution corresponds to F-type dwarfs
and subgiants with a mean absolute magnitude $M_{Ks}=2.5^m$. The shift of this maximum toward large
$(J-Ks)$ with increasing $Ks$ reflects the reddening of these stars with increasing heliocentric distance. The
distribution of the sample of stars over $Ks$, $l$, and $b$ cells with a statistically significant number of stars in
each cell corresponds to their distribution over three-dimensional spatial cells. As a result, the reddening
$E_{(J-Ks)}$ has been determined with an accuracy of $0.03^m$ for spatial cells with a side of 100 pc. All of
the known large absorbing clouds within 1600 pc of the Sun have manifested themselves in the results
obtained. The distances to the near and far edges of the clouds have been determined with a relative
accuracy of 15\%. The cases where unknown clouds are hidden behind known ones on the same line of
sight have been found. The distance dependence of reddening is considered for various Galactic latitudes
and longitudes. The absorbing matter of the Gould Belt is shown to manifest itself at latitudes up to 40$^{\circ}$ and
within 600 pc of the Sun. The size and influence of the Gould Belt may have been underestimated thus far.
The absorbing matter at latitudes up to 60$^{\circ}$ and within 1600 pc of the Sun has been found to be distributed
predominantly in the first and second quadrants in the southern hemisphere and in the third and fourth
quadrants in the northern hemisphere. The warping of the absorbing layer in the near Galaxy apparently
manifests itself in this way. A nonrandom orientation of the clouds relative to the Sun is possible. The
mass of the baryonic dark matter in solar neighborhoods can then be considerably larger than is generally
believed.

\newpage
\section*{THE METHOD}

The 2MASS catalog (Skrutskie et al. 2006) contains infrared (IR) photometry in the $J$, $H$, and $Ks$
bands for about 470 million stars over the entire sky. More than 70 million 2MASS stars, almost all stars
from $Ks=5^{m}$ to $Ks=13^{m}$ and a significant fraction of stars with $Ks=13-14^{m}$, have an accuracy of their
photometry in all three bands better than $0.05^{m}$. In this study, we consider the reddening of these stars.

The $(J-Ks)$ -- $Ks$ diagrams for 2MASS stars have a characteristic appearance presented in Fig. 1 for
three typical regions of the celestial sphere. The reddening of stars and interstellar extinction increase
from the first region to the third one. In Fig. 1a, three regions of an enhanced distribution density of stars
from the Galactic equator. These are also seen on the contour map of the density distribution of stars shown
in Fig. 2a (the white toning indicates a high distribution density). Here, the $Ks$ magnitude is replaced
with $R=10^{(Ks+2.5)/5}$,which, as is shown below, may be considered as the distance for the stars of some
group. As our analysis showed, the left region of an enhanced distribution density (``white'' stars) contains
mostly early-type main-sequence (MS) stars and subgiants, on average, of spectral type F (then O-F stars),
the middle region (``yellow'' stars) contains mostly red giant branch (RGB) and red giant
clump (RGC) stars, thick-disk giants (TDG) and subgiants, and the right region (``red'' stars) contains
mostly MS stars of spectral type M.

In addition to the listed main groups of stars in Fig. 1a, there are a few bluest and reddest stars,
white and brown dwarfs, respectively, among faint stars. The stars of spectral types O, B, A (OBA)
with $J-Ks<0.2^{m}$ near the Galactic plane, which are almost absent in Fig. 1a, are comparatively numerous
in Fig. 1b. In Fig. 1c, Galactic bulge giants dominate among the most strongly reddened stars. These are
also noticeable in Fig. 1b. 

In Figs. 1b and 2b the maxima of the distribution of stars are shifted toward large $(J-Ks)$ compared
to those in Figs. 1a and 2a due to the reddening of stars. Faint stars reddened more strongly. This
reflects the distance--magnitude correlation for any group of stars with a small spread in absolute magnitudes $M_{Ks}$
and with a sufficiently uniform distribution in space.

These features of the $(J-Ks)$ -- $Ks$ diagrams for the 2MASS catalog were used by Lopez-Corredoira
et al. (2002), Drimmel et al. (2003), and Marshall et al. (2006) to develop the method for selecting
stars, determining their distances, and constructing three-dimensional stellar reddening and interstellar
extinction maps considered here. The presumed RGC stars were selected in these studies. However, as
was shown by Marshall et al. (2006) and can be seen from Fig. 1, the RGC stars near the Sun are
mixed on the diagram under consideration with more distant higher-luminosity RGB stars at middle and
high Galactic latitudes and with nearby dwarfs and subgiants (Figs. 1b and 2b) or bulge stars (Figs. 1c
and 2c) near the Galactic equator due to the high reddening. In addition, as we showed previously
(Gontcharov 2008, 2009a), the RGC stars in the Galactic disk do not form a homogeneous group.
These include RGC stars proper with completely different metallicities and ages from $10^{8}$ to $10^{10}$ yr and
a significant admixture of RGB stars with the same color. Therefore, for many Galactic regions, including
the nearest 2 kpc, the methods for selecting RGC stars on the $(J-Ks)$ -- $Ks$ diagram applied so far do
not free the sample from a noticeable admixture of extraneous stars. As a result, the derived distances,
reddenings, and extinctions can differ significantly from the true ones. This was shown by Drimmel
et al. (2003) and Marshall et al. (2006) when comparing various extinction maps. These maps agree
with one another only for $|b|<10^{\circ}$, $-90^{\circ}<l<+90^{\circ}$
and only at a heliocentric distance from 2 to 8 kpc. In fact, all of the extinction maps obtained from the
photometry of giants from 2MASS either do not contain any data for the region within 2 kpc of the
Sun or these data are unreliable.

In this study, we use O-F stars instead of giants to construct a three-dimensional reddening map for
stars within 1600 pc of the Sun.

Let us outline the essence of the method. If the composition of the group of stars (e.g., O-F stars)
under consideration does not change with increasing $Ks$ (or changes predictably), i.e., the mean absolute
magnitude $\overline{M_{Ks}}$ corresponding to the maximum of the distribution density of group stars does not depend
(or depends predictably) on $Ks$, then the mean distance $\overline{R}$ to the stars of this group corresponds to
the mean magnitude $\overline{Ks}$:
\begin{equation}
\label{lgr}
\log~\overline{R}=(\overline{Ks}-\overline{M_{Ks}}+5-\overline{A_{Ks}})/5,
\end{equation}
where $\overline{A_{Ks}}$ is the mean interstellar extinction in the
$Ks$ band for the group of stars. Within several kpc of the Sun, outside the direction toward the Galactic
center, $\overline{A_{Ks}}$ is negligible compared to $\overline{Ks}$ and $\overline{M_{Ks}}$.
Therefore, Eq. (1) can be rewritten as:
\begin{equation}
\label{rk}
\log~\overline{R}=\overline{Ks}/5+f(Ks),
\end{equation}
where $f(Ks)$ is a predictable function of $Ks$, which is
a small correction.

Dividing the group of stars into $Ks$ cells with a statistically significant number of stars in each cell
and applying the correction $f(Ks)$, we divide the group into three-dimensional spatial cells.
The mean color index of the stars $(J-Ks)$ is calculated for each cell. The difference
\begin{equation}
\label{ejk}
E_{(J-Ks)}=\overline{(J-Ks)}-\overline{(J-Ks)_{0}}
\end{equation}
is considered as the mean reddening of the stars from this group in a given spatial cell. The mean color index
of the unreddened stars from this group $\overline{(J-Ks)_{0}}$ is determined from nearby stars. As a result, we may
consider the dependence of $E_{(J-Ks)}$ on Galactic spherical ($l$, $b$, $R$) or rectangular ($X$, $Y$, $Z$) coordinates.

O-F stars are quite suitable for this method, because they possess the following properties:

(1) have a luminosity high enough to be observed far from the Sun, though nearer than giants;

(2) have a space density high enough for the number of stars in each spatial cell under consideration to
be sufficient for statistical inferences;

(3) form a fairly homogeneous group with a small spread in $M_{Ks}$ and a predictable dependence of $\overline{M_{Ks}}$
and $\overline{(J-Ks)_{0}}$ on $Ks$ (i.e., eventually on $f(Ks)$), because
the admixtures of the remaining classes of stars
in the region of space under consideration and in the
range of $Ks$ and $(J-Ks)$ under consideration are small
and well known.

\section*{THE ACCURACY AND LIMITATIONS OF THE METHOD}

Using Monte Carlo simulations, we will determine to what extent the suggested method is justified. For
this purpose, we will reproduce the composition of the 2MASS catalog, primarily O-F stars, and will
take into account the influence of reddening on the distribution of stars on the $(J-Ks)$ -- $Ks$ diagram. In
addition, we will estimate the accuracy of the distances obtained by this method.

The normal and uniform distributions of quantities realized by the Microsoft Excel 2007 random number
generator, whose general description was given by Wichman and Hill (1982), were used in these simulations.
We generated 200 thousand model stars.

These simulations are largely based on the Besancon model of the Galaxy (BMG) (Robin et al. 2003)
as applied to the solar neighborhood with a radius of 2 kpc and with allowance made for the above
actual 2MASS limitation $Ks<14^{m}$ with regard to the stars with accurate photometry. This limitation, in
fact, excludes white dwarfs from our consideration. In addition, there are negligibly few halo and bulge stars
in this region of space, while the density of thin- and thick-disk stars may be considered to be dependent
only on the Galactic coordinate $Z$. This simplifies considerably the simulations.

Our simulations confirmed the assumption that the RGC, RGB, TDG, subgiants, and dwarfs are
mixed on the $(J-Ks)$ -- $Ks$ diagram to produce a complex dependence of $\overline{(J-Ks)_{0}}$ on $Ks$. Therefore,
detailed simulations were continued only for O-F stars, i.e., stars with $(J-Ks)_{0}<0.5^{m}$.

Let us specify the initial mass function according to the BMG: we calculate the stellar mass
(in solar masses) as $m=(m')^{-1/\alpha}$, where $m'$ is a random variable distributed uniformly in the interval
from 0.001 to 20 and $\alpha$ is a parameter that depends on the stellar mass in the BMG. The mass dependence
of $\alpha$ reflects the observed features of the initial mass function. In accordance with the BMG, we take
$\alpha=0.5$ for thick-disk stars and $\alpha=1.6$ at $m<1~M_{\odot}$ and $\alpha=3.0$0 at $m>1~M_{\odot}$ in the thin disk.

We assume the star formation in the Galactic disk to be constant over the last $10^{10}$ yr. This determines
the distribution of stars in age and metallicity. Small changes in disk age have no effect on the results. No
computations for a variable star formation rate have been performed so far, because these are longer at
least by an order of magnitude. Such computations are planned in future. Allowance for the variability
of star formation will apparently change the three-dimensional reddening map only slightly. This follows
from the fact that the corrections obtained in our simulations and used in this study were confirmed in
analyzing real data, which is discussed below.

We specify the lifetime of each model star and the time of its stay at various evolutionary stages from the
MS to a white dwarf as a function of the stellar mass and metallicity in accordance with the evolutionary
tracks and isochrones by Girardi et al. (2000). The same tracks and the properties of the photometric $J$
and $Ks$ bands determine $(J-Ks)_{0}$ and $M_{Ks}$ for each model star.

We specify a uniform distribution of stars in rectangular Galactic coordinates $X$ and $Y$ and a normal
distribution in $Z$ with an age-dependent dispersion $\sigma(Z)=140T^{0.4}$ pc, where $T$ is the age
in Gyr. This distribution of stars in $Z$ agrees with an acceptable accuracy with the observed one presented
by Veltz et al. (2008). Under the influence of the spiral pattern, the OBA stars can be distributed nonuniformly
in space. However, since the absorbing matter also follows the spiral pattern and since the latitude
dependence of extinction is much stronger than its longitude dependence, it makes sense to search for
a systematic dependence of the results on longitude instead of simulating a nonuniform distribution of
stars in longitude.

We calculate the Galactic spherical coordinates of the model stars $R$, $l$, $b$ and the interstellar extinction
dependent on them $A_V$ from the rectangular coordinates by taking into account the extinction in
the Gould Belt based on a new analytical extinction model (Gontcharov 2009b).

The dependence of extinction $A_{\lambda}$ on the effective wavelength $\lambda$ is considered in many studies (see,
e.g., Nishiyama et al. 2006, Marshall et al. 2006, and references therein). Having analyzed these data
by taking into account the effective wavelengths $\lambda_{V}=0.553$ ìêì and $\lambda_{Ks}=2.16$ microns, in this study we
adopted
\begin{equation}
\label{avak}
A_{V}=9.09A_{Ks}=5.88E_{(J-Ks)}.
\end{equation}
Given $A_V$, we calculate $A_{Ks}$ and $E_{(J-Ks)}$ from this formula. Subsequently, we calculate the reddening distorted
color index $(J-Ks)=(J-Ks)_{0}+E_{(J-Ks)}$. Next, we calculate the magnitude of each model star
$Ks=M_{Ks}-5+5\log(R)+A_{Ks}$ and leave only the stars with $Ks<14^{m}$. Analyzing their distribution on
the $(J-Ks)$ -- $Ks$ diagram, we calculate the dependence of $\overline{M_{Ks}}$ and $\overline{(J-Ks)_{0}}$ on $Ks$
as well as the deviations of the reddening $E_{(J-Ks)}$ calculated from
Eq. (3) and the distance $\overline{R}$ calculated from Eq. (1) from the true ones.

Our simulations lead us to the following main conclusions confirmed by our analysis of real data.

(1) Evolved F-type stars (subgiants) account for an appreciable fraction of the O-F stars under consideration
and, filling a narrow range of $(J-Ks)_{0}$ and $M_{Ks}$ (the bluer stars have already become giants,
while the redder ones have not year left the MS) and showing no dependence of $(J-Ks)_{0}$ on $Ks$, they
almost remove this dependence even for the entire set of O-F stars.

(2) The admixture of OBA stars in sky regions near the Galactic equator acts to reduce $\overline{(J-Ks)_{0}}$. As
a result of our simulations, we adopted an empirical correction to $\overline{(J-Ks)_{0}}$ dependent on the latitude $b$:
\begin{equation}
\label{delta1}
\Delta\overline{(J-Ks)_{0}}=-0.09\sin(b).
\end{equation}
This correction reduces all $\overline{(J-Ks)_{0}}$ to the value of $0.23^{m}$ obtained for $b=0$ in our simulations.
The same value of $\overline{(J-Ks)_{0}}$ was obtained for nearby unreddened O-F stars near the Galactic plane from the
Hipparcos catalog (ESA 1997).

(3) For O-F stars, we obtained $\overline{M_{Ks}}=2.5^{m}$. For the mentioned Hipparcos stars, we obtained the same
value (i.e., on average, these are F-type dwarfs and subgiants).

(4) O-F stars are not mixed with giants and dwarfs of spectral type M for $Ks<13.5^{m}$ and $\overline{(J-Ks)}<0.8^{m}$.
These values were confirmed in analyzing real data. Given $\overline{(J-Ks)_{0}}=0.23^{m}$, we find from
Eq. (3) that the suggested method is applicable for $E_{(J-Ks)}<0.57^{m}$, i.e., given Eq. (4), for $A_{V}<3.3^{m}$.
The absorbing medium corresponds to this limitation at least within 1300 pc of the Sun toward the Galactic
center and within 1600 pc of the Sun in the remaining directions (Gontcharov 2009b).

(5) Given $\overline{M_{Ks}}\approx2.5$ and $Ks<13.5^{m}$, we calculate
the limiting distance to which the method in question is applicable from Eq. (1) using O-F stars:
$R<1600$ pc.

(6) The admixture of subdwarfs far from the Galactic equator for $Ks>13.5^{m}$ leads to a decrease
in $\overline{(J-Ks)_{0}}$. Since the space distribution density of subdwarfs and its dependence on $Z$ are known
poorly, this effect cannot yet be taken into account, while the results of this study referring to $Ks>13.5^{m}$
and, accordingly, to $R>1600$ pc, given the sample incompleteness and the mixing of stars, can
have systematic errors. However, the influence of OBA stars and subdwarfs on the parameters of
the sample of O-F stars is much weaker than the influence of the mixing of various classes of stars with
$(J-Ks)_{0}>0.5^{m}$ in analyzing the sample of giants in the mentioned studies of other authors.

(7) An approximately linear growth of $M_{Ks}$ with $(J-Ks)_{0}$ known for the MS leads to an increase in
$\overline{(J-Ks)_{0}}$ with $Ks$. As a result of our simulations, we adopted an empirical correction
to $\overline{(J-Ks)_{0}}$:
\begin{equation}
\label{delta2}
\Delta\overline{(J-Ks)_{0}}=-0.00004R,
\end{equation}
where $R$ is the distance in pc. The value of this correction was confirmed in studying O-F stars toward
the Galactic poles, which is discussed below. Corrections (5) and (6) are the only quantities taken in
the study from our simulations. In other respects, our results do not depend on the adopted model of the
Galaxy and the theory of stellar evolution.

(8) Given the corrections, $\overline{(J-Ks)_{0}}$ is determined with an accuracy of $0.02^{m}$. Taking into account the
errors in the 2MASS photometry, the deviation of $E_{(J-Ks)}$ determined by the suggested method from
the true one does not exceed $0.03^{m}$ (this corresponds to the determination error $\sigma(A_{V})=0.18^{m}$ (see
Eq. (4)). Taking into account the MS peculiarities for $0.2^{m}<(J-Ks)_{0}<0.4^{m}$, the accuracy of determining
$\overline{M_{Ks}}$ can then be estimated as $0.3^{m}$. Consequently,
since $\overline{Ks}$ and $\overline{A_{Ks}}$ are accurate compared to $\overline{M_{Ks}}$, we
find that, according to Eq. (1). the relative accuracy of determining $\overline{R}$ is 15\%.

The suggested method is applicable only in the case of a sufficiently large number of stars in the
spatial cells under consideration. The mean reddening of the stars is calculated for each cell. Therefore, if
the size of the absorbing cloud is slightly smaller than the cell size, then the cloud will manifest itself as a
cell with a reddening higher than that in neighboring cells. However, in this case, the reddening in the cloud
will be underestimated. If, alternatively, the cloud size is much smaller than the cell size, then the cloud will
not manifest itself in any way.

For a spatial cell slightly larger than the absorbing cloud contained in it, the stars located farther than
the cloud but outside its projection onto the celestial sphere are observed in the part unoccupied by the
cloud. Being outside the cloud, they generally have a lower reddening than the stars in and behind the
cloud. Therefore, the $R$ -- $E_{(J-Ks)}$ relation for such a cell has a maximum corresponding to the distance to
the far cloud edge and, then, at larger $R$, a minimum corresponding to the stars farther from the cloud but
outside its projection onto the celestial sphere. This minimum by no means implies that the reddening
behind the cloud is lower than that in it. Such minima are seen in Figs. 3--6, although theoretically the
reddening on the same line of sight cannot decrease with distance. In fact, this effect is the result of the
spatial cell under consideration being too large in size. Although this effect in the present study, to some
extent, even help localize the absorbing clouds, it is desirable to reduce the spatial cell size to the point of
disappearance of this effect.

Thus, it is desirable that the spatial cells under consideration contain a sufficient number of stars but
do not exceed the sought-for absorbing clouds in size. At the first stage of our investigation, we consider
cells $100\times100\times100$ pc in size. More complex techniques at the next stages of our investigation will
allow the resolution of the method to be increased.

The suggested method gives a low accuracy in determining
the distances to clouds nearer than 200 pc. This is attributable to relatively large star density fluctuations,
inaccurate 2MASS photometry for bright stars and/or selection in favor of stars with highly
accurate photometry, and other factors. Thus, three different approaches in producing extinction maps
are efficient at different heliocentric distances. Nearer than 200 pc, the reddening and extinction measurements
for individual stars are irreplaceable; in the ranges from 200 pc to 2 kpc and from 2 kpc to 8 kpc,
it is better to apply the method under consideration using O-F stars and giants, respectively.

Table 1 presents the reddening $E_{(J-Ks)}$ expressed in hundredths of a magnitude and averaged
for cubes with a side of 100 pc on a coordinate grid from $-1550$ to $1550$ pc for $X$ and $Y$ and from $-900$ to
$900$ pc for Z.

\section*{RESULTS FOR THE LARGEST ABSORBING CLOUDS}

The Galactic spherical coordinates ($l$, $b$, $R$) and angular sizes ($\Delta~l$, $\Delta~b$)
of the largest known absorbing clouds within 1.6 kpc of the Sun from Dame et al. (1987) and Dutra and Bica (2002)
are given in Tables 2 and 3 for the clouds near and far from 
the Galactic plane, respectively. Here, $R_{ref}$ is the distance to the cloud from these studies and $R$ is the
distance determined by the method suggested in this study. Obviously, these lists are incomplete primarily
because unknown far clouds are shielded by known near ones.

The distance $R_{ref}$ is usually determined from a comparison of the cloud's radial velocity with a
Galactic rotation model by assuming that the velocity is determined only by rotation and there are no peculiar
velocities and radial motions. The accuracy of $R_{ref}$ is very low, possibly, no higher than 500 pc in some
cases (Gomez 2006). The distance is determined to the observed part of the cloud, while the predominant
shield part of the cloud may not be observed. The main advantage of the suggested method is that it
gives the distance $R$ to the near and far edges of the cloud and to all of the hidden clouds on the same
line of sight. The position of the cloud is determined to within the spatial cell size, i.e., without applying
the complex techniques to be considered below, with an accuracy of 100 pc. Obviously, this accuracy is
usually considerably higher than that obtained from a comparison of the cloud's radial velocity with a
Galactic rotation model.

Figure 3 shows contour maps of reddening $E_{(J-Ks)}$ as a function of coordinates $X$ and $Y$ in
the layers $Z=+50\div+150$ pc, $Z=-50\div+50$ pc, and $Z=-150\div-50$ pc. The Galactic center is
to the right, the Sun or its projection onto the Galactic plane is at the map center. For the Sun,
$E_{(J-Ks)}=0$. The tonal gradations and contour lines mark $E_{(J-Ks)}$ from $0^{m}$ to $0.56^{m}$ at $0.04^{m}$
steps in the case of Z$=-50\div+50$ and from $0^{m}$ to $0.24^{m}$ at $0.02^{m}$ steps in the remaining cases. The
reddening-producing matter is located at the places of maximum $E_{(J-Ks)}$ gradient, i.e., at the places
where the contour lines in the figure are crowded. The numbers mark approximate positions of the
absorbing clouds from Tables 2 and 3. It is important that all of the 19 clouds under consideration not
only were revealed by the suggested method but also turned out to be approximately at the distances that
had been assigned to them so far. In fact, in all cases, the suggested method confirmed the assumption that
the classical method of comparing the radial velocity with a Galactic rotation model gives only an estimate
of the distance to the nearest edge of the cloud. We see that for many clouds, especially those near the
Galactic plane, there is one or more farther clouds on the same line of sight with a known cloud. This
was also assumed in previous interstellar extinction studies. For example, Bochkarev and Sitnik (1985)
and Straizys et al. (1999) pointed out that the Cygnus Rift apparently extends to a distance from 500 to
2000 pc and has a radial orientation relative to the Sun and an elongated shape with a size ratio
of 1 to 5. The same authors also pointed out that the preferential orientation of dust grains and whole giant
clouds with sizes of hundreds of parsecs could be caused by the Galactic magnetic field. In addition, the
radial orientation of the absorbing clouds within the nearest several hundred parsecs relative to the Sun
can be explained by the Sun's special location not far from the epicenter of the processes that have formed
the Local Bubble and the Gould Belt comparatively recently. As was noted in these papers, the X-ray
emission generated in these processes could affect the chemical composition and, hence, the absorptive
properties of the dust grains. Consequently, the total mass of the light-absorbing matter in Galactic solar
neighborhoods and, hence, the baryonic dark matter may be several times higher than is usually calculated
from the apparent frontal size of clouds by assuming their random orientation relative to the Sun.

It is also important that the suggested method revealed no new absorbing clouds within 1 kpc of
the Sun. This leads us to conclude that the method does not give rise to artifacts. At the same time, the
suggested method reveals several hitherto unknown absorbing clouds farther than 1 kpc from the Sun,
although these, along with the known clouds, belong to the largest structures in Galactic solar neighborhoods,
the Gould Belt and the Local Spiral Arm. More specifically, there are two largest complexes of
clouds well above the Galactic plane: at low Galactic longitudes and in the second Galactic quadrant.
The Cepheus (no. 10) and $\rho$~Oph (no. 12) clouds, respectively, are known at the near edge of these complexes.
The complex toward the Galactic anticenter is located well below the Galactic plane. A whole group
of known clouds is located at its near edge: Perseus (no. 13), Taurus (no. 14), Orion (nos. 15, 16, and 17).

Figure 4 shows the rise in stellar reddening with distance for the clouds from Table 3 (solid lines) in
comparison with the directions closest to them at the same latitude (dashed lines). The clouds are marked
by shading at distances where the rise in $(J-Ks)$ is at a maximum. We see that several clouds lie on the same
line of sight in almost all cases. It is also noticeable that $(J-Ks)<0.23^{m}$ at $R=50$ pc for all clouds. This
means that, in contrast to other directions, there is an appreciable number of unreddened OBA stars within
100 pc in these directions.

\section*{THE DISTANCE DEPENDENCE OF REDDENING FOR VARIOUS LATITUDES}

Figure 5 shows the distance dependence of reddening found for various Galactic latitudes. The
results were averaged in $10^{\circ}$-wide bands in latitude and in regions near the north and south Galactic
poles (NGP, SGP), $b>+80^{\circ}$ and $b<-80^{\circ}$, respectively. In the figure, the dependences are indicated
for each hemisphere by the following curves: the solid heavy curve for the region near the pole, the
heavy dotted curve for $70^{\circ}<|b|<80^{\circ}$, the solid black
curve for $60^{\circ}<|b|<70^{\circ}$, the dash-dotted curve for
$50^{\circ}<|b|<60^{\circ}$, the gray dotted curve for $40^{\circ}<|b|<50^{\circ}$,
and the solid gray curve for $30^{\circ}<|b|<40^{\circ}$. The trends show the mean reddening for the regions near
the poles.

The following conclusions can be reached about the reddening of stars at middle and high latitudes.

Given Eq. (4), the values of $E_{(J-Ks)}$ obtained agree well with known extinction maps and models
at middle and high latitudes.

The mean reddening $\overline{E_{(J-Ks)}}$ near the NGP ($0.025^{m}$) is higher than that near the SGP ($0.01^{m}$). The
reddening $E_{(J-Ks)}$ near the NGP at $R<250$ pc is higher than that at the remaining northern latitudes
under consideration. At all northern latitudes, except the NGP, $E_{(J-Ks)}$ grows with $R$ and latitude more
slowly than it does at the corresponding southern latitudes.
As a result, at $b=-30^{\circ}\div-60^{\circ}$, $E_{(J-Ks)}$ is higher than that at $b=+30^{\circ}\div+60^{\circ}$.
This apparently means that there is only one compact absorbing cloud toward the NGP in the northern hemisphere
within $R<250$ pc, while $E_{(J-Ks)}$ in the southern hemisphere within $R<250$ pc increases at all latitudes
rather rapidly, i.e., southward, the absorbing medium is denser but less clumpy. Thus, the conclusion
by Gontcharov (2009b) that the Sun is located below the central plane of the equatorial absorbing
layer should be corrected. The equatorial absorbing layer in solar neighborhoods apparently cannot be
considered as a flat one. It has a rise within 250 pc of the Sun. However, outside this region, it is located,
on average, below the Sun, in fact, coinciding with the Galactic equator, which lies 13 pc below the Sun
(Gontcharov 2008).

Corrections (5) and (6) found from our simulations reduced $E_{(J-Ks)}$ near the Sun to $0^{m}\pm0.02^{m}$
(which is within the limits of the random errors of the method) and completely removed the growth of
$E_{(J-Ks)}$ with $R$ toward the poles. Thus, the observations completely agree with the simulations.

The reddening $E_{(J-Ks)}<0^{m}$ at $b=+60^{\circ}\div+80^{\circ}$ and $R<250$ pc. This is due to the large number
of unreddened OBA stars, including the stars of the Ursa Majoris cluster closest to the Sun.

Figure 6 shows contour maps of the dependence of $E_{(J-Ks)}$ on $l$ and $R$ found at middle latitudes.
The results were averaged in $10^{\circ}$-wide bands with the centers at $|b|=25^{\circ}$, $35^{\circ}$, $45^{\circ}$
è $55^{\circ}$. The contour lines mark $E_{(J-Ks)}$ from $0^{m}$ at $0.03^{m}$ steps.
The white toning denotes the maximum reddening. The reddening-producing matter lies in the regions
of maximum growth in $E_{(J-Ks)}$, i.e., where the contour lines are crowded.

In the northern and southern hemispheres, the reddening is at a maximum at opposite longitudes.
For example, at $|b|<40^{\circ}$, the absorbing matter is associated mainly with the Gould Belt, which is oriented
in such a way that the absorbing matter lies predominantly above the equator at $l\approx0^{\circ}$ and below
the equator at $l\approx180^{\circ}$ (Gontcharov 2009b). We see
that the corresponding reddening grows up to $R\approx600$ pc. Thus, the Gould Belt affects the distribution
of absorbing matter at distances and latitudes outside
its universally accepted size (300--400 pc, $15-20^{\circ}$).
The size of the Belt, as a region of concentration of
absorbing matter, may be much larger than is generally
believed.

At $|b|>40^{\circ}$, the reddening is caused mainly not by the Gould Belt. It is at a maximum in the first
and second quadrants in the southern hemisphere and in the third and fourth quadrants in the northern
hemisphere. This effect is also present at $|b|<40^{\circ}$, but it is barely noticeable against the background of
reddening in the Gould Belt. This is apparently how the warping of the Galactic absorbing layer manifests
itself, not on the scales of the Galaxy but in some part of it: this warping is opposite to than found by
Marshall et al. (2006) for regions farther than 3 kpc from the Sun.

\section*{CONCLUSIONS}

This is the second paper in our series of studies
of the interstellar extinction in the Galaxy. The previous
study (Gontcharov 2009b) showed the necessity
of allowance for the extinction in the Gould Belt in
constructing an analytical model and the shortage
of accurate extinction measurements in the nearest
kiloparsec. This study is a step toward the solution
of this problem: a detailed accurate three-dimensional
extinction map can serve as a material for analytical
models and analysis of Galactic structures. In addition,
by assigning the mean reddenings to stars in
accordance with the three-dimensional map, it will
allow the determination of their basic characteristics
from multicolor photometry to be simplified.

Currently available large-scale catalogs, for example,
the 2MASS catalog with accurate photometry
for 70 million stars used in this study, are not
only the main material for present-day studies of
the Galaxy but also a ``testing ground'' for developing
the methods of automatically determining the
fundamental characteristics of stars from multicolor
photometry in future space projects. The method of
determining the distances and reddenings of stars
from $(J-Ks)$ -- $Ks$ diagrams under consideration is
among the most promising methods, because it uses
only infrared photometry in two broad bands.

Monte Carlo simulations and analysis of real data
confirmed the conclusion reached by the previous
researchers who applied this method: using giants
allows an extinction map to be constructed only at
a heliocentric distance from 2 to 8 kpc. Nearer than
2 kpc, this method is worth applying to F-type dwarfs
and subgiants. This narrows the region of space under
consideration but allows a three-dimensional reddening
map for stars to be constructed with an accuracy
$\sigma(E_{(J-Ks)})=0.03^{m}$ (which corresponds to
$\sigma(A_{V})=0.18^{m}$) on a spatial grid with a cell of no
more than 100 pc within no less than 1600 pc of the
Sun. This allowed us to detect the largest absorbing
clouds and to determine the distances to them
with a relative accuracy of 15\%. In some cases, more
distant clouds manifested themselves behind known
ones on the same line of sight. Most of the clouds
detected outside the Galactic plane fit into the Gould
Belt, whose sizes and significance may have been
underestimated so far. Everywhere, except the Gould
Belt, the distribution of absorbing matter, especially
farther than 500 pc, is apparently determined by the
warping of its layer in such a way that its density is
higher in the first and second Galactic quadrants in
the southern hemisphere and in the third and fourth
quadrants in the northern hemisphere.

The detection of absorbing clouds and the determination
of their distances are particularly important
in connection with their possible nonrandom orientation
in solar neighborhoods and in the entire Galaxy.
If other clouds are hidden behind known ones on the
same lines of sight or the clouds obey the Galactic
magnetic field or the special position of the Sun
relative to nearby clouds, the Local Bubble, and the
Gould Belt has the same cause, then constructing
a three-dimensional map of absorbing matter is of
great importance in estimating the amount and role of
baryonic dark matter and constructing cosmological
models.

\section*{ACKNOWLEDGMENTS}

In this study, we used results from the Hipparcos
and 2MASS (Two Micron All-Sky Survey) projects as
well as resources from the Strasbourg Astronomical
Data Center (France). This study was supported by
the Russian Foundation for Basic Research (project
no. 08-02-00400) and in part by the ``Origin and
Evolution of Stars and Galaxies'' Program of the
Presidium of the Russian Academy of Sciences.

\newpage

\begin{figure}[p]
\includegraphics{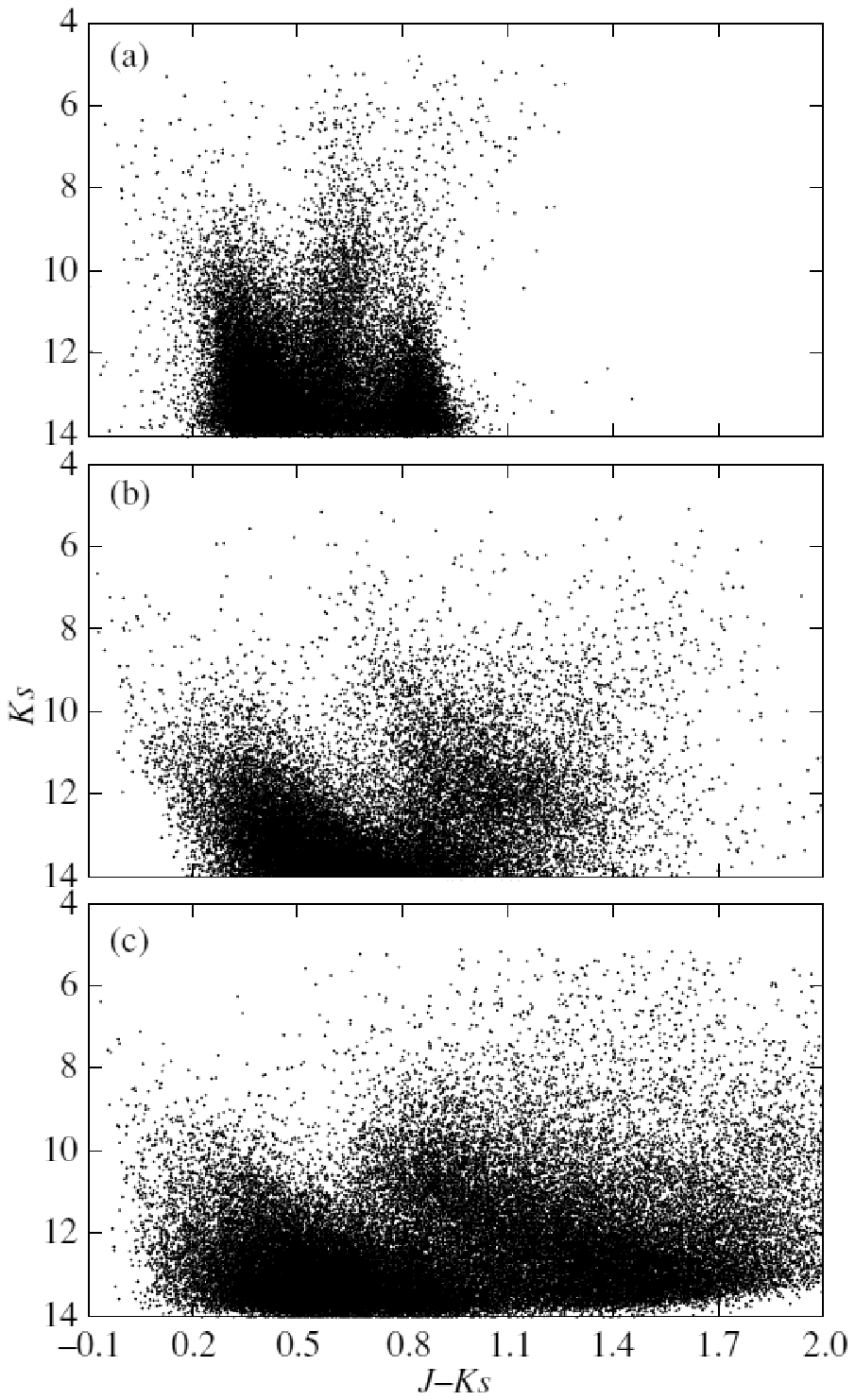}
\caption{$(J-Ks)$ -- $Ks$ diagram for 2MASS stars with accurate
photometry: (a) near $l=90^{\circ}, b=+30^{\circ}$; (b) near
$l=180^{\circ}, b=0^{\circ}$; (c) near $l=90^{\circ}, b=0^{\circ}$.
}
\label{jkk}
\end{figure}

\newpage

\begin{figure}[p]
\includegraphics{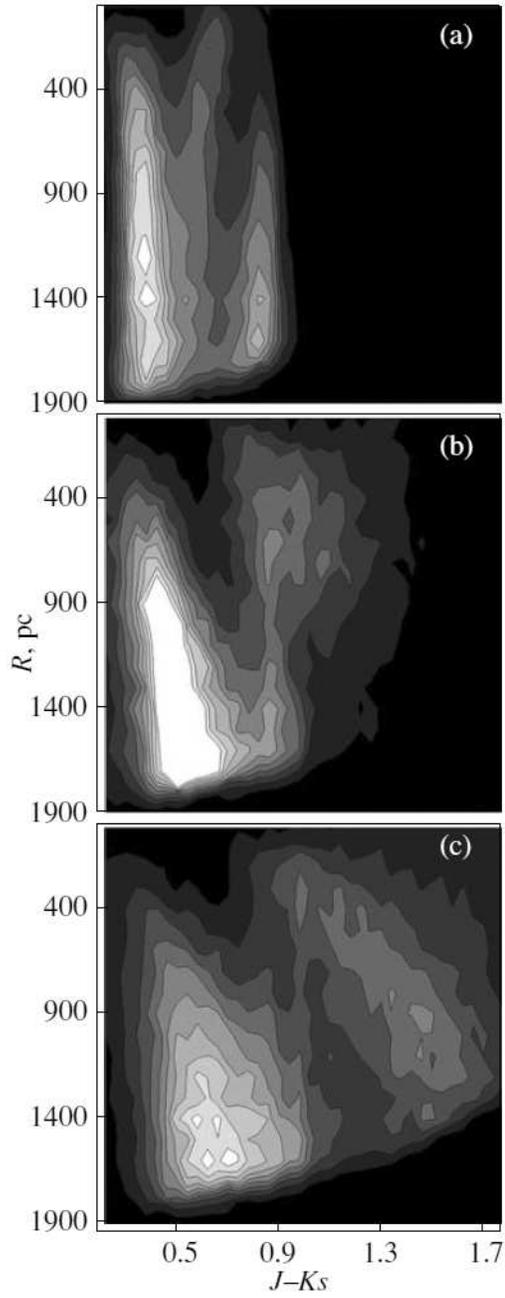}
\caption{Contour map of the density distribution of 2MASS
stars with accurate photometry on the $(J-Ks)$ -- $R$ diagram,
where $R=10^{(Ks+2.5)/5}$: (a) near $l=90^{\circ}, b=+30^{\circ}$; (b) near $l=180^{\circ}, b=0^{\circ}$;
(c) near $l=90^{\circ}, b=0^{\circ}$.
}
\label{jkkmap}
\end{figure}

\newpage

\begin{figure}[p]
\includegraphics{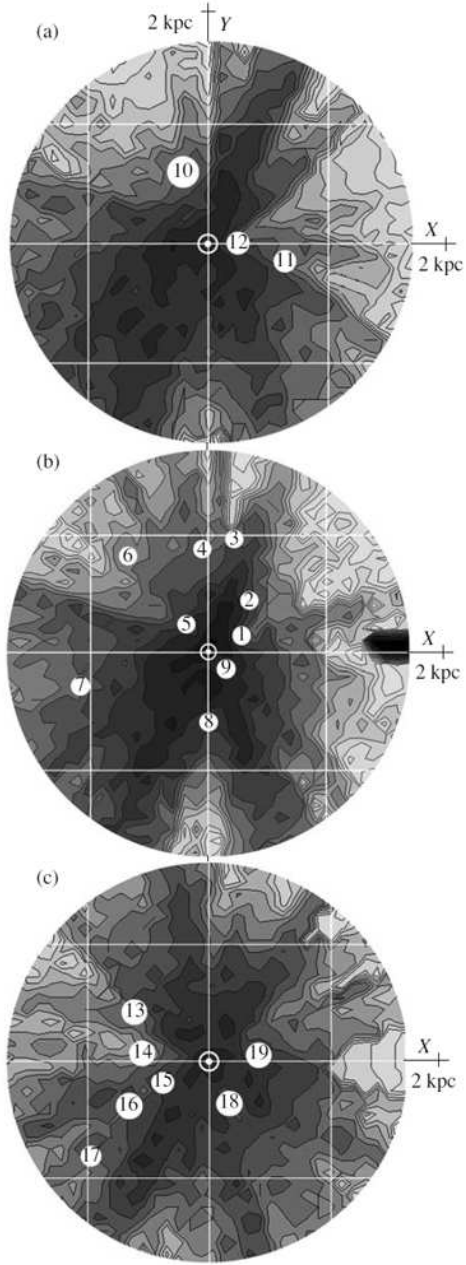}
\caption{Contour maps of reddening $E_{(J-Ks)}$ as a function of coordinates $X$ and $Y$ in the following layers:
(a) $Z=+50\div+150$, (b) $Z=-50\div+50$, and (c) $Z=-50\div-150$ pc.
The white lines of the coordinate grid are plotted at 1-kpc steps. The numbers mark the positions of the largest known absorbing clouds from Tables 2 and 3.
}
\label{exy}
\end{figure}

\newpage

\begin{figure}[p]
\includegraphics{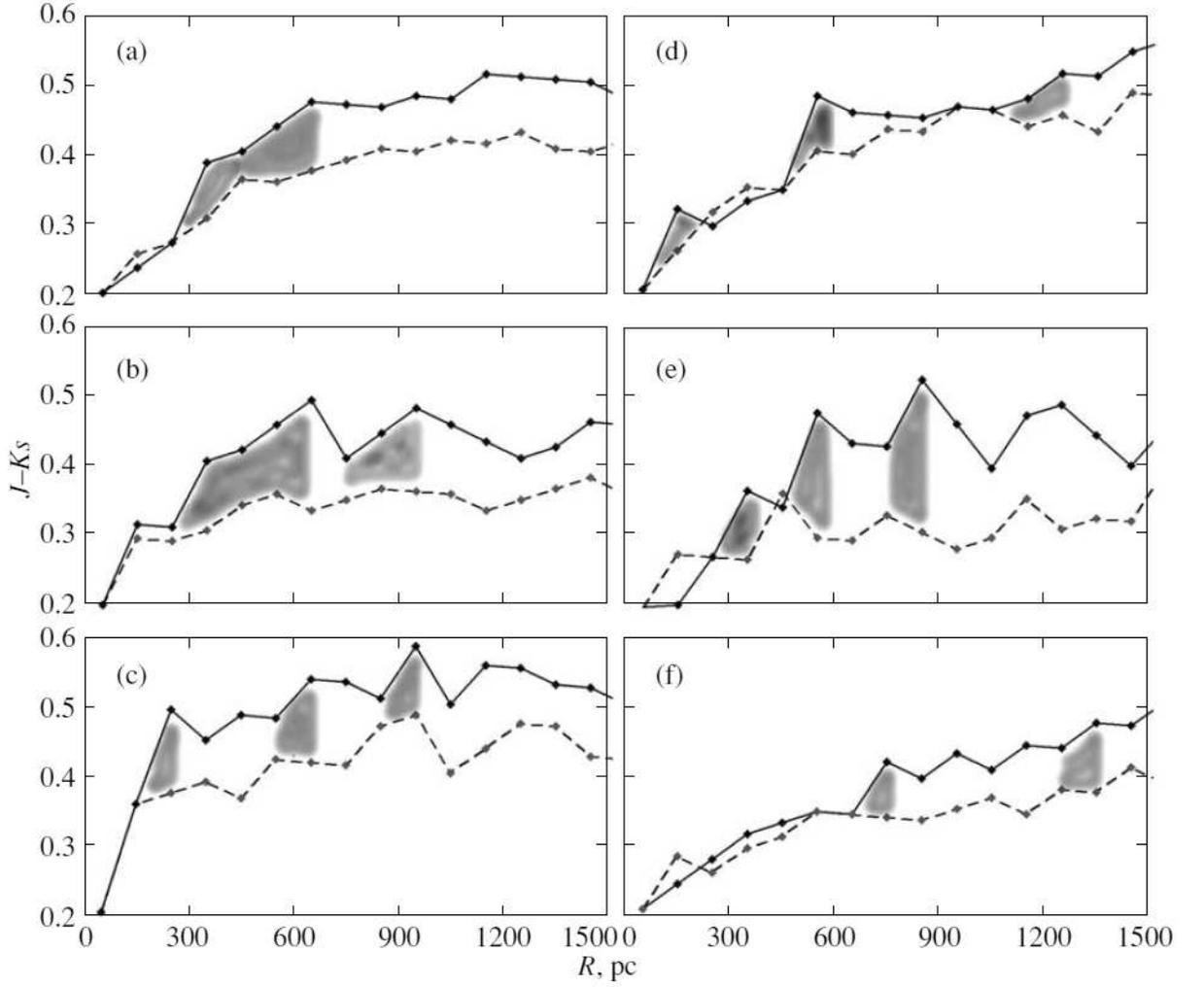}
\caption{$J-Ks$ color index corresponding to the maximum of the distribution of O-F stars versus heliocentric distance
for the largest clouds outside the Galactic plane:
(a) Cepheus, (b) Perseus, (c) Taurus, (d) Orion~B, (e) Orion~A, (f) Mon~R2.
The solid and dashed lines indicate the dependence for the cloud and for the region closest to the cloud at the same
latitude, respectively.
The cloud positions are marked by shading.
}
\label{clouds}
\end{figure}

\newpage

\begin{figure}[p]
\includegraphics{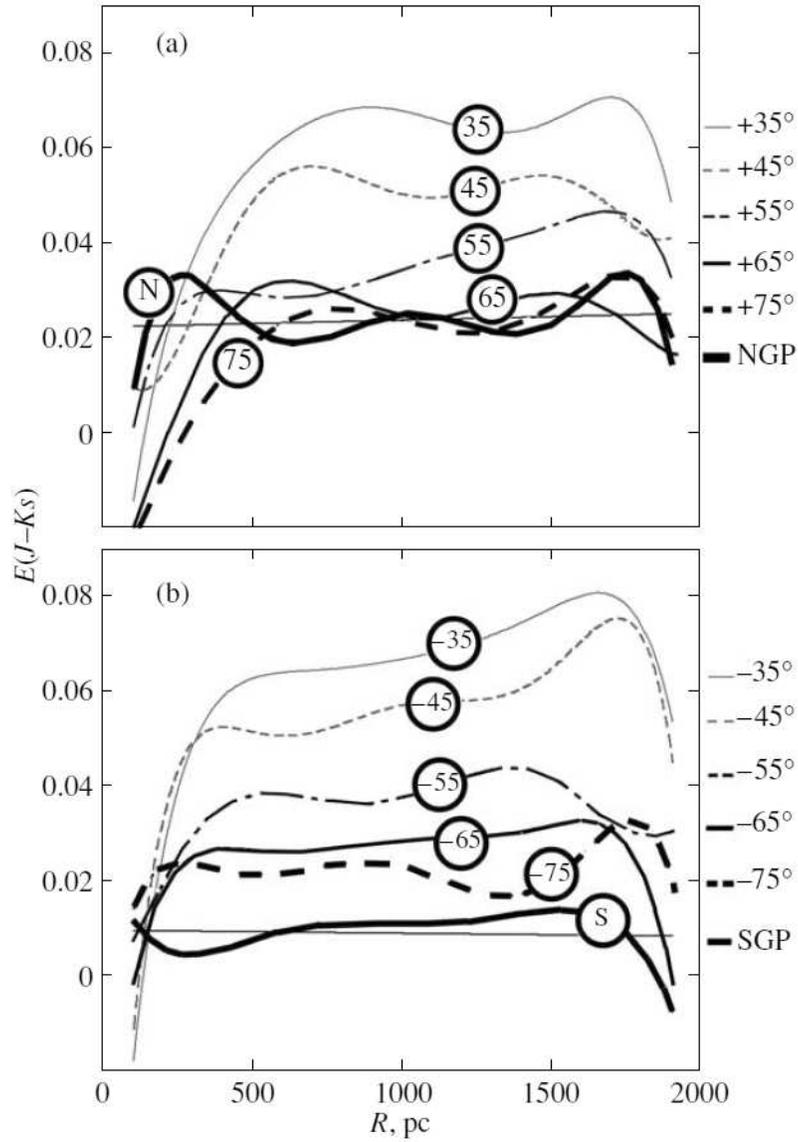}
\caption{Reddening $E_{(J-Ks)}$ versus distance for various Galactic latitudes.
}
\label{high}
\end{figure}

\newpage

\begin{figure}[p]
\includegraphics{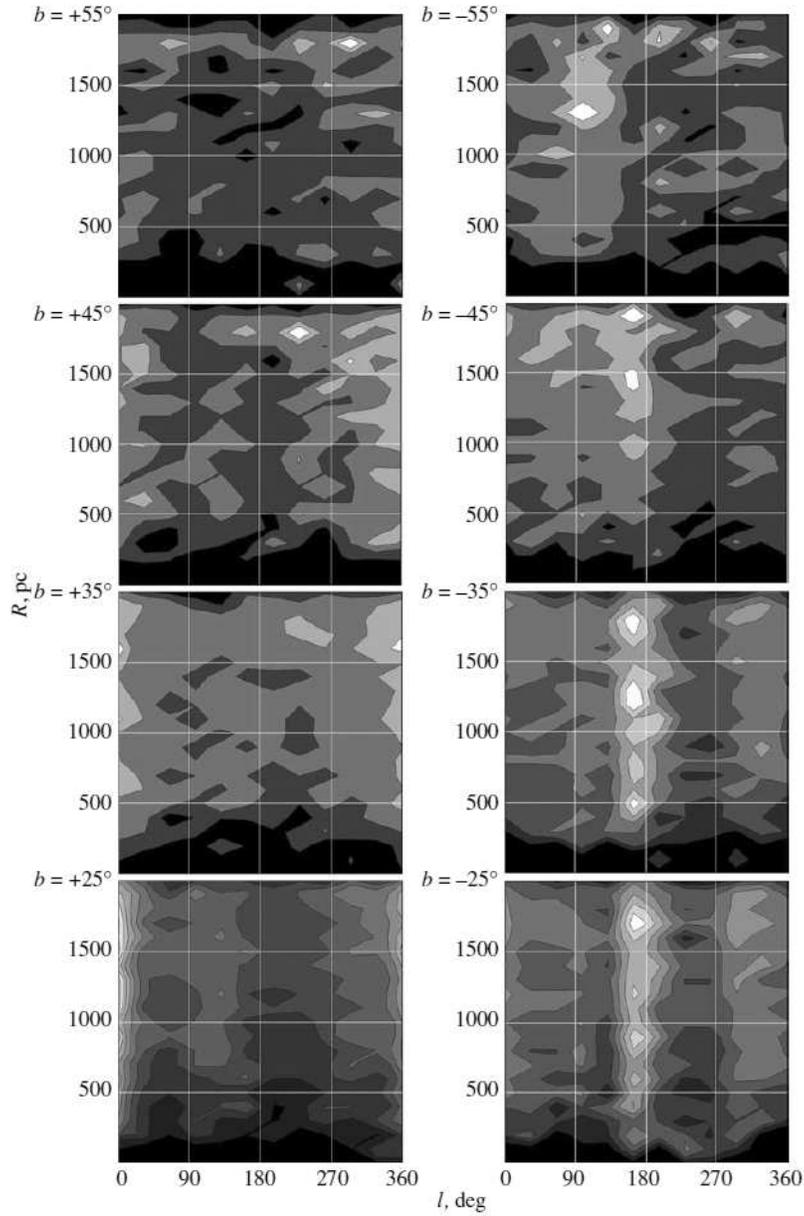}
\caption{Contour maps of reddening $E_{(J-Ks)}$ as a function of distance and Galactic longitude for 10$^{\circ}$
bands in Galactic latitude. The contour lines mark $E_{(J-Ks)}$ from $0^{m}$ at $0.03^{m}$ steps.
}
\label{midlat}
\end{figure}

\newpage

\begin{table*}[!h]
\def\baselinestretch{1}\normalsize\footnotesize
\caption[]{Reddening $E_{(J-Ks)}$ expressed in $0.01^{m}$ as a function of coordinates $X$, $Y$, and $Z$
averaged for cubes with a side of 100 pc.
Part of the data for the layer $-50<Z<+50$ pc is given, the table is entirely available in electronic form
(There was an erratum here -- it is corrected in arXiv -- as pointed out by Gontcharov (2012, Astronomy Letters, 38, 87):
in all readings on the $Y$ scale, the sign should be reversed, for example, $+550$ and $-550$ pc
should be replaced by $-550$ and $+550$ pc, respectively.)
}
\label{xyze}
\[
\begin{tabular}{rrrrrrrrrrrrr}
\hline
\noalign{\smallskip}
\multicolumn{4}{l}{$-50<Z<+50$}  & & & & & & & & & \\
\hline
$Y/X$, pc & $-550$ & $-450$ & $-350$ & $-250$ & $-150$ & $-50$ & 50 & 150 & 250 & 350 & 450 & 550 \\
\hline
\noalign{\smallskip}
$-550$  & 09 & 07 & 05 & 08 & 08 & 10 & 10 & 08 & 10 & 15 & 11 & 17 \\
$-450$  & 07 & 05 & 08 & 06 & 08 & 04 & 08 & 08 & 10 & 08 & 13 & 17 \\
$-350$  & 07 & 06 & 08 & 04 & 08 & 05 & 05 & 08 & 08 & 12 & 14 & 13 \\
$-250$  & 08 & 06 & 06 & 07 & 05 & 03 & 07 & 09 & 11 & 12 & 14 & 16 \\
$-150$  & 08 & 08 & 04 & 05 & 07 & 07 & 09 & 09 & 09 & 14 & 16 & 18 \\
$-50$   & 10 & 08 & 09 & 07 & 01 & 03 & 03 & 05 & 13 & 13 & 14 & 18 \\
$+50$  & 14 & 12 & 13 & 09 & 05 & 02 & 03 & 11 & 17 & 21 & 14 & 26 \\
$+150$ & 12 & 14 & 14 & 11 & 11 & 05 & 03 & 09 & 17 & 26 & 16 & 20 \\
$+250$ & 22 & 24 & 18 & 15 & 11 & 09 & 05 & 07 & 13 & 18 & 26 & 28 \\
$+350$ & 25 & 26 & 16 & 14 & 14 & 11 & 05 & 04 & 12 & 16 & 26 & 29 \\
$+450$ & 21 & 19 & 16 & 18 & 14 & 08 & 10 & 08 & 14 & 18 & 19 & 25 \\
$+550$ & 21 & 19 & 19 & 20 & 18 & 14 & 10 & 08 & 12 & 19 & 25 & 27 \\
\hline
\end{tabular}
\]
\end{table*}

\newpage

\begin{table*}[!h]
\def\baselinestretch{1}\normalsize\footnotesize
\caption[]{Coordinates and sizes of the largest known absorbing clouds near the Galactic plane
}
\label{clouds0}
\[
\begin{tabular}{rlrrrrcc}
\hline
\noalign{\smallskip}
N & Name  & $l$, deg & $b$, deg & $\Delta~l$, deg & $\Delta~b$, deg & $R_{ref}$, pc & $R$, pc  \\
\hline
\noalign{\smallskip}
1 & Aquila rift                      &  30 &   2 &  26  &  10 &  200       &  200--800  \\
2 & Vulpecula rift                   &  59 &   1 &   9  &   8 &  400       &  300--1500 \\
3 & Cygnus rift                      &  75 &   0 &  24  &   8 &  700       &  700--1200 \\
4 & Cyg OB7 CO complex               &  93 &   3 &  12  &  11 &  800       &  400--1600 \\
5 & Lindblad ring                    & 132 &   3 &  64  &  14 &  300       &  300--1000 \\
6 & $-$12 km/s CO clouds             & 131 &   3 &  59  &  14 &  800       &  800--1000 \\
7 & Mon OB1 + Rosette complex        & 204 &   0 &   8  &   3 &  800--1600  &  600--1600 \\
8 & Vela complex                     & 270 &   2 &  17  &   7 &  420--500   &  400--1500 \\
9 & Coal Sack + Lower Scorpius  & 324 &   0 &  50  &   4 &  150       &  150--1500 \\
\hline
\end{tabular}
\]
\end{table*}

\newpage

\begin{table*}[!h]
\def\baselinestretch{1}\normalsize\footnotesize
\caption[]{Coordinates and sizes of the largest known absorbing clouds far from the Galactic plane
}
\label{clouds1}
\[
\begin{tabular}{rlrrrrcc}
\hline
\noalign{\smallskip}
N & Name  & $l$, deg & $b$, deg & $\Delta~l$, deg & $\Delta~b$, deg & $R_{ref}$, pc & $R$, pc  \\
\hline
\noalign{\smallskip}
10 & Cepheus            & 110 & $+$17 & 20 & 11 & 450 & 350--650  \\
11 & Lupus              & 340 & $+$13 & 13 & 18 & 170 & $>150$ \\
12 & $\rho$~Oph         & 354 & $+$18 & 12 & 11 & 111 & $>100$ \\
13 & Perseus            & 160 & $-$18 &  8 &  8 & 350 & 350--650, 750--950 \\
14 & Taurus             & 172 & $-$14 & 15 & 12 & 350 & 250, 650, 950 \\
15 & Orion B            & 206 & $-$14 &  5 & 10 & 450 & 150, 550, 1250 \\
16 & Orion A            & 211 & $-$20 &  9 &  6 & 500 & 350, 550, 850 \\
17 & Mon~R2 + Crossbones & 215 & $-$11 &  8 &  4 & 830 & 750, 1350 \\
18 & Chamaeleon         & 300 & $-$16 & 10 &  8 & 215 & 150--350 \\
19 & R~CrA              &   0 & $-$18 &  7 &  8 & 150 & 150 \\
\hline
\end{tabular}
\]
\end{table*}


\begin{thebibliography}{99}

\bibitem{bosi} N.G.~Bochkarev, T.G.~Sitnik, Astron. Astrophys. Suppl. Ser. \textbf{108}, 237 (1985).

\bibitem{dame} T.M.~Dame, H.~Ungerechts, R.S.~Cohen, \emph{et al.}, Astrophys. J. \textbf{322}, 706 (1987).

\bibitem{dcl} R.~Drimmel, A.~Cabrera-Lavers, M.~Lopez-Corredoira, Astron. Astrophys. \textbf{409}, 205 (2003).

\bibitem{dutra} C.M.~Dutra, E.~Bica, Astron. Astrophys. \textbf{383}, 631 (2002).

\bibitem{hip} ÅÊÀ (ESA), \emph{Hipparcos and Tycho catalogues} (ESA, 1997).

\bibitem{g2000} L.~Girardi, A.~Bressan, G.~Bertelli, \emph{et al.}, Astron. Astrophys. Suppl. Ser. \textbf{141}, 371 (2000).

\bibitem{gomez} G.C.~Gomez, Astron. J. \textbf{132}, 2376 (2006).

\bibitem{rcg} G.A. Gontcharov, Pis'ma Astron. Zh. \textbf{34}, 868 (2008) [Astron. Lett. \textbf{34}, 785 (2008)].

\bibitem{monte} G.A. Gontcharov, Pis'ma Astron. Zh. \textbf{35}, 707 (2009a) [Astron. Lett. \textbf{35}, 638 (2009a)].

\bibitem{gould} G.A. Gontcharov, Pis'ma Astron. Zh. \textbf{35}, 862 (2009b) [Astron. Lett. \textbf{35}, 780 (2009b)].

\bibitem{lc} M.~Lopez-Corredoira, A.~Cabrera-Lavers, F.~Garzon, \emph{et al.}, Astron. Astrophys. \textbf{394}, 883 (2002).

\bibitem{marsh} D.J.~Marshall, A.C.~Robin, C.~Reyle, \emph{et al.}, Astron. Astrophys. \textbf{453}, 635 (2006).

\bibitem{nishi} S.~Nishiyama, T.~Nagata, N.~Kusakabe, \emph{et al.}, Astrophys. J. \textbf{638}, 839 (2006).

\bibitem{robin} A.C.~Robin, C.~Reyle, S.~Derriere, \emph{et al.}, Astron. Astrophys. \textbf{409}, 523 (2003).

\bibitem{2mass} M.F.~Skrutskie, R.M.~Cutri, R.~Stiening, \emph{et al.}, Astron. J. \textbf{131}, 1163 (2006),
http://www.ipac.caltech.edu/2mass/releases/allsky/index.html.

\bibitem{strai} V.~Straizys, C.J.~Corbally, V.~Laugalys, Baltic astronomy, 8, 355 (1999)

\bibitem{veltz} L.~Veltz, O.~Bienayme, K.C.~Freeman, \emph{et al.}, Astron. Astrophys. \textbf{480}, 753 (2008).

\bibitem{random1} B.A.~Wichman, I.D.~Hill, Applied Statistics \textbf{31}, 188 (1982).


\end{thebibliography}
\end{document}